\documentclass[conference,10pt]{IEEEtran}
\usepackage[english]{babel}
\usepackage[T1]{fontenc}
\usepackage{cite,url,color} 
\usepackage{graphics,amsfonts}
\usepackage{epstopdf}
\usepackage[pdftex]{graphicx}
\usepackage[cmex10]{amsmath}
\usepackage[utf8]{inputenc}

\usepackage{enumitem}

\usepackage{tikz}
\usetikzlibrary{automata,positioning,chains,shapes,arrows}
\usepackage{pgfplots}
\usetikzlibrary{plotmarks}
\newlength\fheight
\newlength\fwidth
\pgfplotsset{compat=newest}
\pgfplotsset{plot coordinates/math parser=false}

\usepackage{array}
\usepackage{subfig}

\usepackage[top=1.5cm, bottom=2cm, right=1.6cm,left=1.6cm]{geometry}
\usepackage{indentfirst}

\usepackage{times}

\usepackage{breqn}
\usepackage{booktabs}

\usepackage{placeins}

\begin{document}
\title{M2M Massive Access in LTE: RACH Performance Evaluation in a Smart City Scenario}

\author{\IEEEauthorblockN{Michele Polese,
				  Marco Centenaro,
				  Andrea Zanella,
				  and Michele Zorzi}
\IEEEauthorblockA{Department of Information Engineering, University of Padova -- Via Gradenigo, 6/b, 35131 Padova, Italy\\Email: {\tt\{polesemi,marco.centenaro,zanella,zorzi\}@dei.unipd.it}
}}

\begin{verbatim}
This article has been accepted for publication in the Proceedings of 
the 2016 IEEE International Conference on Communications (ICC).

PLEASE, CITE THE PAPER AS FOLLOWS:

Plain text:
M. Polese, M. Centenaro, A. Zanella, and M. Zorzi, M2M Massive Access in LTE: 
RACH Performance Evaluation in a Smart City Scenario, 2016 IEEE International Conference 
on Communications (ICC), Kuala Lumpur, 2016, pp. 1-6.

BibTex:
@INPROCEEDINGS{7511430, 
author={M. Polese and M. Centenaro and A. Zanella and M. Zorzi}, 
booktitle={2016 IEEE International Conference on Communications (ICC)}, 
title={M2M massive access in LTE: RACH performance evaluation in a Smart City scenario}, 
year={2016}, 
pages={1-6}, 
keywords={Long Term Evolution;multi-access systems;public domain software;
smart cities;Internet of Things;LTE Random Access Channel;
LTE cellular standard;LTE module;LTE random access procedure;
Long Term Evolution cellular standard;M2M massive access;RACH;
Smart City scenario;mobile terminals;open-source network simulators;
system-level simulators;Delays;Indexes;Internet of things;
Long Term Evolution;Signal to noise ratio;Smart cities;Uplink}, 
doi={10.1109/ICC.2016.7511430}, 
month={May},}

\end{verbatim}

\maketitle

\begin{abstract}
Several studies assert that the random access procedure of the Long Term Evolution (LTE) cellular standard may not be effective whenever a massive number of simultaneous connection attempts are performed by terminals, as may happen in a typical Internet of Things or Smart City scenario.
Nevertheless, simulation studies in real deployment scenarios are missing because many system-level simulators do not implement the LTE random access procedure in detail.
In this paper, we propose a patch for the LTE module of {\em ns--3}, one of the most prominent open-source network simulators, to improve the accuracy of the routine that simulates the LTE Random Access Channel (RACH).
The patched version of the random access procedure is compared with the default one and the issues arising from massive simultaneous access from mobile terminals in LTE are assessed via a simulation campaign.
\end{abstract}
\begin{picture}(0,0)(0,-320)
\put(0,0){
\put(0,0){\footnotesize This paper was accepted for presentation at the IEEE ICC 2016} 
\put(0,-10){\footnotesize conference, May 23 - 27, 2016, Kuala Lumpur, Malaysia.}}
\end{picture}

\section{Introduction}\label{sec:intro}
A huge increase in fully automated communications between devices is forecasted in the next few years as new use cases, e.g., connected cars, e-health, environmental monitoring, are being identified. 
This new paradigm is called Machine-to-Machine (M2M) communication, since it is typically performed without any human intervention, and is considered a fundamental enabler of the Internet of Things (IoT) vision.

A desirable requirement for the deployment of Machine-Type Devices (MTDs) is the \textit{place \& play} concept \cite{biral}, i.e., MTDs should just need to be deployed in a certain area to be ready to operate.
Indeed, the expected number of devices (40 MTDs per household according to \cite{tr45820}) makes a manual configuration infeasible.
For this reason, the cellular network infrastructure is suitable to provide connectivity for M2M communications, since it can provision (ideally) worldwide, ubiquitous coverage, in contrast to many ad hoc proprietary technologies.
However, deploying such a huge number of devices in current cellular networks, e.g., the Long Term Evolution (LTE), poses new issues that need to be addressed.
In particular, an important problem is the overload of the LTE Random Access (RA) procedure under a massive number of simultaneous connection attempts, since the LTE standard has been designed to provide high-rate access to a fairly limited number of terminals.

In this paper, we aim at evaluating the delay that a device may undergo while accessing an LTE network in the case of a massive number of access requests in a real deployment.
In particular, we address a Smart City scenario using one of the most accurate open-source system-level network simulators, i.e., \textit{network simulator 3} (ns--3, \cite{ns3}).
We found that the current implementation of the RA procedure in ns--3 is idealized; therefore, we developed a patch to make the routine suitable to study the impact of M2M traffic in LTE networks in urban scenarios.
Simulation results show that if a few hundred smart sensors simultaneously require network access, e.g., to report some kind of failure event, an MTD would experience extremely long delays to complete the access procedure, thus not respecting the delay constraints of important Smart City applications, such as alarms.

The remainder of the paper is organized as follows.
In Sec.~\ref{sec:issues}, the random access procedure in LTE is described and issues related to massive access are briefly explained.
In Sec.~\ref{sec:RACH_in_LTE} the implementation of LTE Random Access Channel (RACH) in ns--3 is discussed and its weaknesses are highlighted; then, a patch to the default routine is proposed to enhance the accuracy of the simulator.
In Sec.~\ref{sec:perf_eval} the patched routine is compared with the default one and is used to evaluate the impact of a massive number of simultaneous access attempts in a realistic Smart Cities scenario.
Finally, conclusions are drawn in Sec.~\ref{sec:conclusion}.

\section{The Random Access Procedure in LTE}\label{sec:issues}
The RA procedure in LTE is initiated when (i) a User Equipment (UE) is in the {\tt RRC_CONNECTED} state\footnote{A UE is in the {\tt RRC_CONNECTED} state when a Radio Resource Control (RRC) connection has been established; if this is not the case, the UE is in the {\tt RRC_IDLE} state \cite[Sec.~4.2.1]{36331}.} and has new data to transmit or receive but no uplink synchronization; (ii) it recovers after radio link failure; (iii) it switches from {\tt RRC_IDLE} to {\tt RRC\_CONNECTED}, or, finally, (iv) it performs a handover.
The procedure takes place in a dedicated physical channel called Physical Random Access Channel (PRACH) \cite{sesia}, which is multiplexed in time and frequency with the Physical Uplink Shared Channel (PUSCH).
The PRACH consists of $6$ Resource Blocks (RBs) for an overall bandwidth of $1.08$~MHz and has a duration between $1$ and $4$ subframes.
Its periodicity is variable and is defined by the PRACH Configuration Index, which is broadcast by the base station (eNodeB, eNB) on the System Information Broadcast 2 (SIB2) along with the following signaling information:
\begin{itemize}
\item {\tt numContentionPreambles}, i.e., the number of preambles reserved for contention-based RA (at most $64$);
\item {\tt preambleInitialReceivedTargetPower}, i.e., the target power (in dBm) to be reached at the eNB for transmissions on PRACH;
\item {\tt powerRampingStep}, i.e., the power ramping step used to increase the transmission power after every failed attempt;
\item {\tt preambleTransMax}, i.e., the maximum number of preamble transmission attempts.
\end{itemize}
We recall that the {\em preamble} is a {\em signature} composed of a cyclic prefix and a Zadoff-Chu (ZC) sequence that is obtained by shifting a {\em root sequence}, which is common to all the UEs connected to a certain eNB.
Preambles containing different sequences are orthogonal to one another.\footnote{For eNBs with a large coverage area there may be more than one root sequence. However the sequences obtained have low cross-correlation \cite{son}.}
There are $4$ different preamble formats, with duration from 1 to 4 subframes, in order to guarantee coverage of different cell sizes. 

The RA procedure consists of $4$~messages as follows.
\paragraph{Preamble Transmission} The UE selects a random ZC sequence and transmits a preamble on one of the resources specified by the PRACH Configuration Index.
The eNB will detect the sequence by applying a correlator and a peak detector to the received signal \cite{sesia}. 
However, since the number of ZC sequences is finite, it may happen that more than one UE select the same sequence, thus incurring in a \emph{collision}. 
If the colliding UE preambles are received with high enough Signal-to-Noise Ratio (SNR), and are sufficiently spaced apart in time, two energy peaks separated by a time that is longer than the Maximum Delay Spread (MDS) are detected and the eNB will interpret this event as due to a collision. 
On the other hand, if only one of the colliding preambles is received with high SNR, or the delay of the different preambles is similar,\footnote{This is typical in Small Cells scenarios \cite{sesia}.} the eNB will not be able to recognize the collision.

We remark that MDS and the preamble detection algorithm are not standardized but left to the eNB vendor.
However, the 3GPP report \cite{3gpp104} requires a missed detection probability lower than $10^{-2}$ for an SNR value of $-14.2$ dB and a 2-antenna receiver in AWGN channel.
\paragraph{Random Access Response (RAR)} The eNB answers to correctly decoded preambles (including those with undetected collision) by sending a RAR message on the Downlink Shared Channel (DLSCH). 
RAR carries the detected {\em preamble index}, which corresponds to the sequence sent by the UE, a timing alignment to synchronize the UE to the eNB, a temporary identifier (RNTI), and an uplink scheduling grant that specifies the resources assigned to the UE to transmit in the next phase of the RA procedure.
If a UE receives a RAR, then it proceeds with the third step; otherwise, it restarts the RA procedure anew (unless it has reached the maximum number of preamble transmission attempts) after a backoff time that is uniformly distributed in the interval $[0, ${\tt BI}$]$, where {\tt BI} is the Backoff Indicator carried by the RAR.
If the counter of consecutive unsuccessful preamble transmissions exceeds the maximum number of attempts, a RA problem is indicated to the upper layers.
\paragraph{Connection Request} The UE transmits a Radio Resource Control (RRC) message containing its core-network terminal identifier in the uplink grant resources and starts a Contention Resolution timer.
Note that the UEs that transmitted the same preamble but whose collision remained undetected will transmit on the same resources, colliding again.
\paragraph{Contention Resolution} If the eNB correctly receives the RRC message, it replies with an RRC Connection Setup that signals to the UE that the RA phase is successfully completed.
Instead, if the Contention Resolution timer expires, the UE repeats the RA procedure from the beginning after a random backoff time.
Again, when the number of unsuccessful attempts reaches some specified maximum value, the network is declared unavailable by the UE and an access problem exception is raised to the upper layers.

\subsection{Massive Random Access}
The LTE RA process is efficient when a small number of devices require access to the network, which is the typical case for Human-to-Human (H2H) traffic.
However, the number of terminals is envisioned to grow exponentially in IoT scenarios, especially for what concerns Smart Cities where {\em smart meters} will be deployed to monitor a large variety of parameters, from air pollution to supply levels.
In the case of extraordinary events, e.g., power outages, many MTDs may be activated simultaneously, thus causing a PRACH overload.
The consequences are that the constraints of delay-sensitive M2M applications can be violated, the power consumption of the sensors is increased, and the Quality of Service (QoS) of H2H applications can be degraded.

In this paper we aim at evaluating how this massive {\em event-triggered} reporting may impact the LTE network performance in a Smart City scenario using a well-known open-source network simulation tool, i.e., ns--3, written in C++, which is particularly suitable to simulate an urban propagation environment.
Other open-source simulation platforms are available, e.g., the LTE Vienna Simulator \cite{viennaSimulator}, which is based on Matlab, and Omnet++ \cite{SimuLTE} and LTE-sim \cite{LTE-sim}, both written in C++.
However, they cannot be directly used for our purposes.
In fact, the Vienna Simulator is a link level simulator for the uplink, and therefore lacks some of the necessary features to adequately model a network of MTDs, whereas Omnet++ and LTE-sim focus on the higher networking layers through an idealized abstraction of the lower layers, and therefore do not capture the level of detail we need to model the RACH performance.


\section{LTE RACH in ns--3}\label{sec:RACH_in_LTE}
\begin{figure}[t]
	\centering
	\includegraphics[width = 0.5\textwidth]{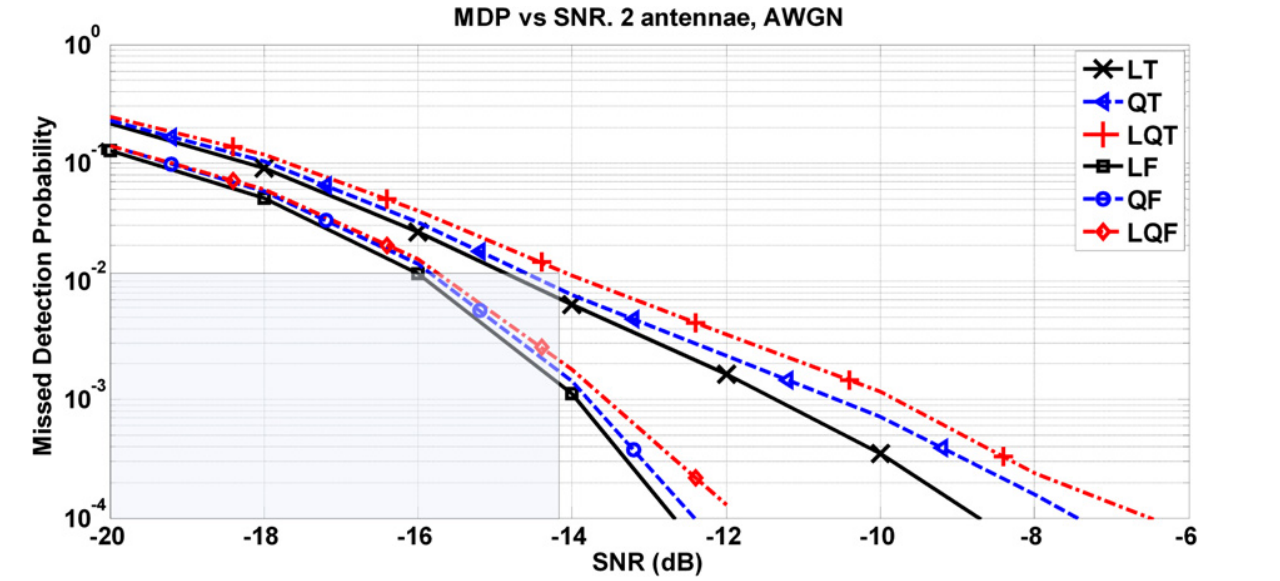}
	\caption{$P_{\rm miss}$ vs SNR $\Gamma$, taken from \cite{perfeval}.}
	\label{fig:pmiss}
\end{figure}

In this section the LTE RACH implementation in ns--3 is addressed and an enhancement to evaluate IoT traffic is proposed.

We refer to version 3.23 of the ns--3 simulator, which uses the LTE-EPC Network simulAtor (LENA) \cite{lena} module to simulate the LTE protocol stack and the Evolved Packet Core (EPC) network.
In the current implementation of LENA, however, the RACH preamble is an {\em ideal} message, i.e., not subject to radio propagation; moreover, Connection Request and Connection Resolution messages are not modeled and, therefore, all collisions are detected and solved at the first step of the RA procedure. 
Furthermore, we found that it is not possible to simulate the connection and disconnection of UEs during runtime, since LENA allows every UE to switch only once from {\tt RRC_IDLE} to {\tt RRC_CONNECTED} states at the beginning of the simulation.
Therefore, we implemented a more realistic RA procedure, along with the possibility to disconnect UEs from the eNB (i.e., switching from {\tt RRC_IDLE} to {\tt RRC_CONNECTED} and vice versa).
The enhanced module is called LENA+.
\footnote{The source code is available at \texttt{\url{https://github.com/signetlabdei/lena-plus}}.} 
However, to maintain the backward compatibility with the current release, an option has been introduced to use the idealized LENA RA procedure if desired.
In the following, we describe in detail the features of LENA+ that were not present in LENA.

\begin{table}[t]
\centering
\begin{tabular}{c|c}
\toprule
Parameter 						&	Value 			\\
\midrule
Downlink carrier frequency 		& 945 MHz			\\
Uplink carrier frequency		& 900 MHz			\\
RB bandwidth					& 180 kHz			\\
Available bandwidth				&  50 RB 			\\
Hexagonal sectors				& 1				\\
eNBs for each sector				& 3 (co-located)	\\
eNBs beamwidth (main lobe)		&  $65^{\circ}$			\\
TX power used by eNBs 			&  43 dBm 			\\
Max TX power used by MTDs		&  23 dBm 			\\
eNB noise figure				&	3 dB 			\\
MTD noise figure 				&	5 dB 			\\
Shadowing						& 	log-normal with $\sigma = 8$ \\
Number of buildings				&	96 				\\
Apartments for each floor		&	6 				\\
Floors for each building		&	3 				\\
MTD speed 						&   0 Km/h 			\\
Number of MTDs	$N$				& \{50, 100, 150, 200, 300, 400, 500, 600\} \\
Simulation time $ \forall N$	& \{60, 60, 120, 120, 300, 300, 400, 400\} s \\
\bottomrule
\end{tabular}
\caption{Simulation parameters \cite{tr45820}.
} 
\label{table:param}
\end{table}

\begin{table}[t]
\centering
\begin{tabular}{c|c}
\toprule
Parameters  							&	Value 			\\
\midrule
PRACH Configuration Index 				& 	1				\\
Backoff Indicator {\tt BI}				&   0 ms 			\\
{\tt preambleInitialReceivedTargetPower}& 	$-110$~dBm		\\
{\tt powerRampingStep} 					&   2 dB 			\\
{\tt numContentionPreambles} 			& 	54				\\
{\tt preambleTransMax} 					& 	$\infty$ 		\\
Contention resolution timer 			& 	32 ms 			\\
\bottomrule
\end{tabular}
\caption{Simulation parameters of LTE RACH.} 
\label{table:rachparam}
\end{table}

\begin{figure}[t]
	\centering
	\includegraphics[scale=.6]{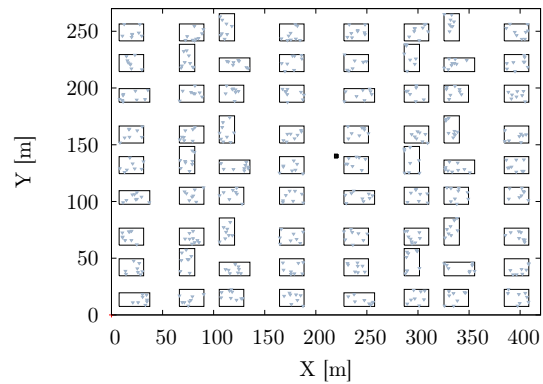}
	\caption{Smart City network deployment example. The rectangles are the buildings, the small triangles are the MTDs, and the black square is the position of the three co-located eNBs.}
	\label{fig:scenario}
\end{figure}

\begin{figure*}[t]
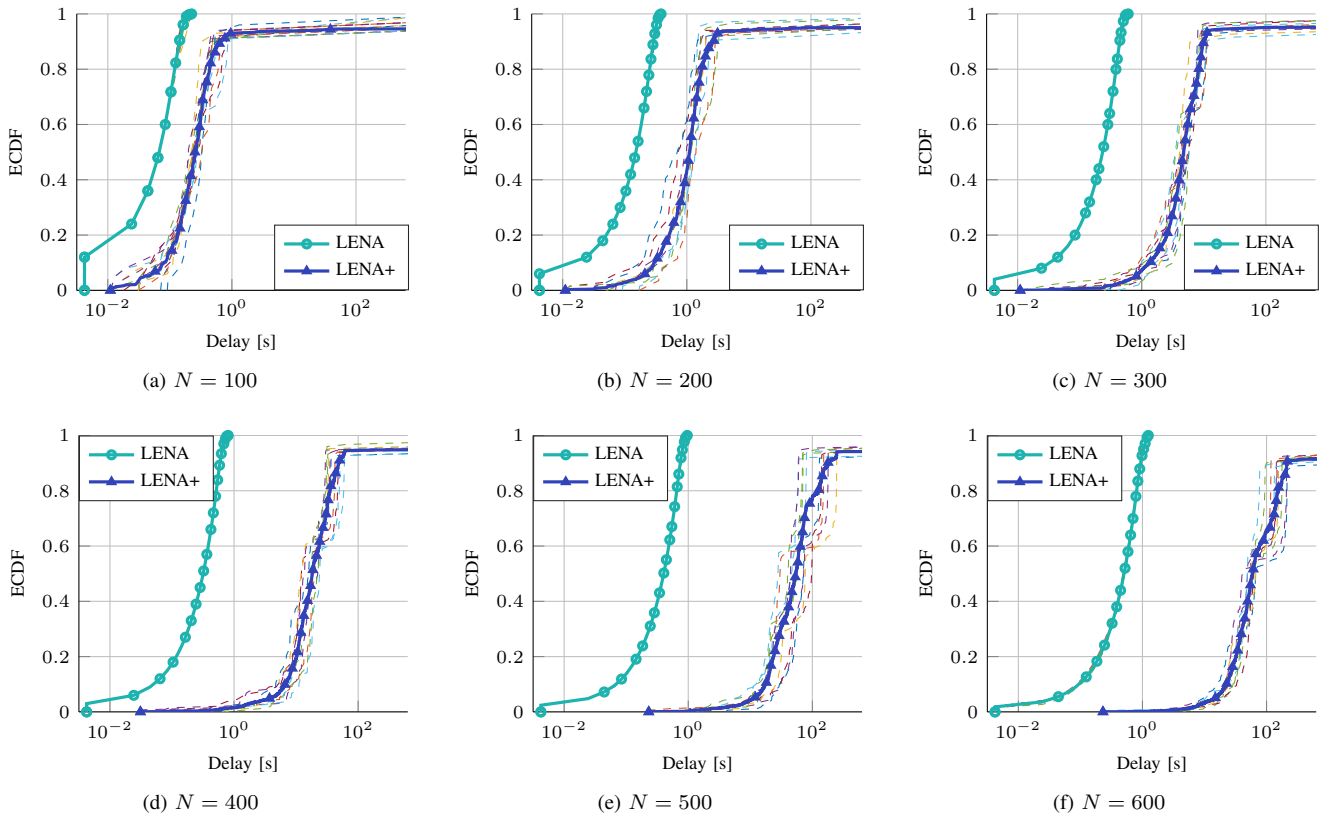

\centering
\subfloat[$N = 100$]{\setlength\fheight{0.2\textwidth}
\setlength\fwidth{0.25\textwidth}
\input{./figures/ecdf_compare_100.tex}
\label{fig:ecdf100}
}
\hfil
\subfloat[$N = 200$]{\setlength\fheight{0.2\textwidth}
\setlength\fwidth{0.25\textwidth}
\input{./figures/ecdf_compare_200.tex}
\label{fig:ecdf200}
}
\hfil
\subfloat[$N = 300$]{\setlength\fheight{0.2\textwidth}
\setlength\fwidth{0.25\textwidth}
\input{./figures/ecdf_compare_300.tex}
\label{fig:ecdf300}
}
\hfil
\subfloat[$N = 400$]{\setlength\fheight{0.2\textwidth}
\setlength\fwidth{0.25\textwidth}
\input{./figures/ecdf_compare_400.tex}
\label{fig:ecdf400}
}
\hfil
\subfloat[$N = 500$]{\setlength\fheight{0.2\textwidth}
\setlength\fwidth{0.25\textwidth}
\input{./figures/ecdf_compare_500.tex}
\label{fig:ecdf500}
}
\hfil
\subfloat[$N = 600$]{\setlength\fheight{0.2\textwidth}
\setlength\fwidth{0.25\textwidth}
\input{./figures/ecdf_compare_600.tex}
\label{fig:ecdf600}
}
\caption{ECDFs of access delay for $N \in \{100,\ 200,\ 300,\ 400,\ 500,\ 600\}$. The x-axis is expressed in logarithmic scale.}
\label{fig:ecdfs}
\end{figure*}

\subsection{PRACH Characterization}
PRACH is implemented as a real physical channel, relying on the already developed and tested channel model.
Nonetheless, PRACH preambles are now subject to noise and radio propagation, since they are sent on specific time and frequency physical resources, and therefore the eNB can fail their detection.
We remark that only format {\tt 0} of PRACH preambles is implemented.
Whenever a UE starts the RA procedure, it checks whether it has received SIB2, which carries the RACH configuration.
Then, it chooses a random index drawn uniformly in [$0$, {\tt numContentionPreambles} - 1] and transmits it in the next PRACH opportunity.
Note that we are assuming that, in a PRACH-dedicated subframe, no PUSCH traffic is allocated by the scheduler, while the PRACH is allocated in the first $6$~RBs available.
The power (in dBm) for the transmission is computed according to the standard as \cite{3gpp213}
\begin{equation}
\begin{aligned}
	&P_{\rm prach} = \min \{ P_{\rm UE, max}, \\ 
		&\quad \texttt{PREAMBLE_RECEIVED_TARGET_POWER} + P_{\rm lc}\}
\end{aligned}
\end{equation}
where $P_{\rm UE, max}$ is the maximum transmit power for a UE, $P_{\rm lc}$ is the estimated pathloss and {\tt PREAMBLE_RECEIVED_TARGET_POWER} is given by the MAC layer, as \cite{3gpp321}
\begin{equation}
\begin{aligned}
	&\texttt{PREAMBLE\_RECEIVED\_TARGET\_POWER} = \\
	&\quad \texttt{preambleInitialReceivedTargetPower} \\
	&\quad + \Delta_{\rm preamble} + (\texttt{PREAMBLE\_TX\_COUNTER} - 1) \\
	&\quad \times \texttt{powerRampingStep}
\end{aligned}
\end{equation}
where {\tt PREAMBLE_TX_COUNTER} is the number of consecutive preamble transmissions and $\Delta_{\rm preamble}=0$ for format {\tt 0}. 
The other parameters are given by the eNB with SIB2, as explained in Sec.~\ref{sec:issues}.
At the eNB side, the SNR is computed for each preamble and a decision on correct or missed detection is made.
In \cite{perfeval} different eNB detection algorithms are introduced and evaluated in terms of missed detection probability vs SNR at the receiver.
The missed detection probability performance of these algorithms is reported in Fig.~\ref{fig:pmiss}.
Note that our improved PRACH model for ns--3 assumes a time domain detector with decimation (denoted with LT in the legend), which is the simplest algorithm which satisfies the 3GPP requirements in \cite{perfeval}, as can be seen from Fig.~\ref{fig:pmiss}.
The ns--3 LTE module has also been modified, in order to handle the reception of multiple signals in the same time and frequency resources.
Indeed, the default implementation raises an exception whenever two or more signals are received in the same time and frequency resources by the eNB, since the MAC scheduler forbids multiple transmissions in physical channels other than PRACH.
In PRACH, indeed, transmissions cannot be scheduled and the ZC sequences allow multiple access by Code Division Multiple Access (CDMA).
If a preamble is correctly received but there are two or more preambles with the same ZC sequence, then the collision is detected or not according to the following heuristic.
Since the PDP of different users is not simulated in a system-level simulator such as ns--3, as a rule of thumb a collision is detected if 
\begin{equation}
	\frac{d_{\rm max} - d_{\rm min}}{c} > T_{\rm chip}\, ,
\end{equation}
where $d_{\rm max}$ and $d_{\rm min}$ are the distances from the eNB of the farthest and closest colliding UE, respectively, $c$ is the speed of light, $T_{\rm chip} = 1/(2B)$, and $B = 1.08$~MHz is the bandwidth of PRACH.

RAR message transmission was already implemented as a message on the DLSCH.
For each not-collided preamble or undetected collision, an uplink grant is allocated by the scheduler and added to the RAR response; the Backoff indicator has also been added.

Connection Request transmission on granted resources was already implemented, as well.
However, since in the default implementation of RACH all the collisions are resolved at the first step, collisions of messages 3 were not handled and the simulator raised an exception.
This exception has been handled as follows.
Firstly, no capture effect has been considered.
Then, if two or more Connection Request messages collide, they are considered as received with errors, triggering an HARQ (layer 2) retransmission until the maximum number of attempts is reached; after that, the RA procedure starts again.
The Contention Resolution timer, that was also not present in LENA, has been added as well. 
\section{Performance Evaluation}\label{sec:perf_eval}
In this section the proposed enhanced version of the LENA module is evaluated and compared with the current release.
Moreover, we will use our module to evaluate the impact of massive simultaneous accesses to an LTE network in a Smart City scenario.

The scenario we simulated is compliant with the specifications in \cite{tr45820}; the main simulation parameters are in Table~\ref{table:param}, while the RACH-related parameters can be found in Table~\ref{table:rachparam}. 
We refer to an urban environment with a high density of tall buildings; the deployment of buildings and MTDs is depicted in Fig.~\ref{fig:scenario}.
For what concerns the radio propagation model, we employed the ns--3 Hybrid Buildings Propagation Loss Model, which exploits different propagation models to account for several factors, such as the positions of the UE and the eNB (both indoor, both outdoor, one indoor and the other outdoor), the external wall penetration loss of different types of buildings (i.e., concrete with windows, concrete without windows, stone blocks, wood), and the internal wall penetration loss.
We remark that all the MTDs have been placed inside the buildings and their positions are not changed during the simulation according to the specifications found in \cite{tr45820}.

\begin{figure}[t]
\centering
\setlength\fheight{0.2\textwidth}
\setlength\fwidth{0.4\textwidth}
\input{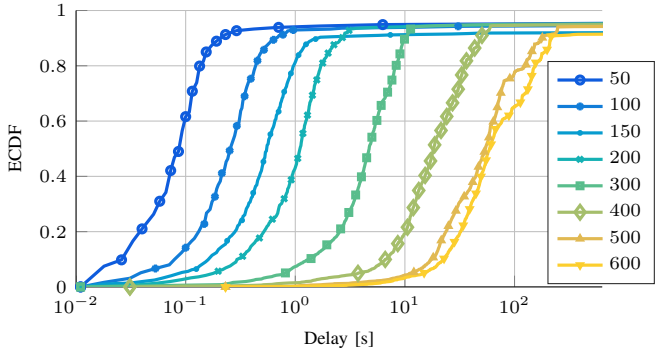}
\caption{ECDF for various values of $N$. The x-axis is expressed in logarithmic scale.}
\label{fig:ecdfall}
\end{figure}

Let us denote with $N$ the number of MTDs that are trying to simultaneously access the LTE network.
For every value of $N \in \{50, 100, 150, 200,300,400,500,600\}$, $10$ Monte Carlo simulations have been run and the Empirical Cumulative Distribution Functions (ECDFs) of the access delays have been produced.
Fig.~\ref{fig:ecdfs} shows the ECDFs of the delay (in logarithmic scale) for the various values of $N$, obtained using both the default LENA module and the LENA+ module.
We remark that, for what concerns the LENA+ performance curves, the average ECDF is represented by the solid line and we plotted the ECDFs of the individual Monte Carlo simulations with dashed lines to show the dispersion around the average value. 
It can be seen that the idealized RACH implemented in LENA gives quite unrealistic results, where all the MTDs would succeed in completing the RA procedure in less than $1$~s for all values of $N$.
The simulations that have been carried out using LENA+, instead, show that, as $N$ grows, the access delay increases, up to hundreds of seconds for most MTDs, which is not acceptable for many delay-constrained Smart City applications, such as alarms.
Moreover, using our module, we are able to observe that some UEs (approximately 5\% of the total in each simulation) do not succeed in completing the RA procedure during the simulation, despite the unlimited number of transmission attempts allowed (i.e., {\tt preambleTransMax} $ = \infty$).
This is due to their unfavorable position, e.g., inside buildings which are far away from the eNB.
An overall comparison among the average ECDFs for all the values of $N$ is provided in Fig.~\ref{fig:ecdfall}, which clearly shows that the access delay increases as $N$ grows.
As a further insight, we invite the reader to refer to Table~\ref{table:means}, which contains the average value and standard deviation of the delays of successful MTDs.
The statistics confirm the trend. 

\begin{table}[t]
\centering
\begin{tabular}{c|c|c|c}
\toprule
$N$ 	&	Mean $\mu$ [s]	& Std. Dev.	$\sigma$ [s]	& $\mu/\sigma$ 			\\
\midrule
50 		& 	0.235	 		& 1.855						& 0.127	\\
100		&   0.498			& 2.608 					& 0.191	\\
150		& 	0.780			& 2.605						& 0.300	\\
200 	&   1.481			& 4.453						& 0.333	\\
300 	& 	5.268			& 5.359						& 0.983	\\
400 	& 	21.400 			& 15.126					& 1.415	\\
500 	& 	64.234			& 52.852					& 1.215	\\
600 	& 	77.423			& 59.256					& 1.307	\\
\bottomrule
\end{tabular}
\caption{Statistics of the access delay experienced by the MTDs that succeeded in completing the access procedure.} 
\label{table:means}
\end{table}

\begin{figure}[t]
\centering
\setlength\fheight{0.2\textwidth}
\setlength\fwidth{0.4\textwidth}
\input{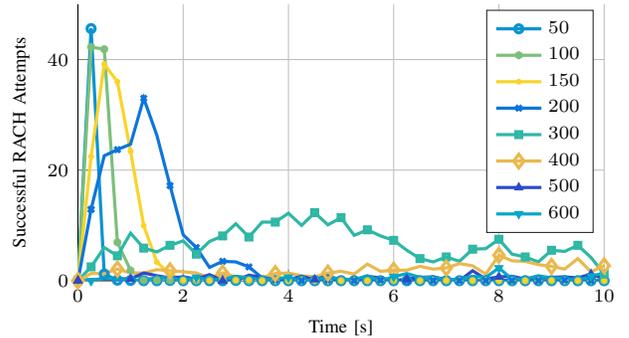}
\caption{Successful RACH attempts vs time.}
\label{fig:goodput}
\end{figure}

Finally, Fig.~\ref{fig:goodput} represents the number of successful RACH attempts vs time for different values of $N$.
Note that, as $N$ increases, the maximum number of successful RACH attempts decreases, and is achieved later in time, as a consequence of the higher number of collision events.
For $N>300$, we cannot observe any meaningful peak, denoting that the RACH is congested and the success probability is very low.


\section{Conclusion}\label{sec:conclusion}
The implementation of a more realistic model of the LTE RA procedure in the network simulator ns--3 has been proposed and evaluated in a Smart City scenario.
An enhanced ns--3 LENA module, called LENA+, has been developed to overcome the limitations of the default routine, which is rather idealized.
The simulation results show that the default release of the LENA module underestimates the impact of M2M traffic in cellular networks and confirms the concerns about LTE RACH overload in case of massive simultaneous access attempts.
As part of our future work, we will enrich the LENA+ implementation, to test the effectiveness of some of the strategies proposed in \cite{biral} to relieve the RACH overload problem.




\begin{thebibliography}{10}

\bibitem{biral}
A. Biral, M. Centenaro, A. Zanella, L. Vangelista, and M. Zorzi, ``The challenges of M2M massive access in wireless cellular networks,'' \textit{Digital Communications and Networks}, Vol. 1, Issue 1, pp. 1-19, Feb. 2015

\bibitem{tr45820}
3GPP, ``Cellular System Support for Ultra Low Complexity and Low Throughput Internet of Things,'' TR 45.820, V. 1.3.1, June 2015

\bibitem{ns3}
\url{https://www.nsnam.org}

\bibitem{36331}
3GPP, ``Radio Resource Control (RRC) -- Protocol specification,'' TS 36.331, V. 12.6.0, June 2015

\bibitem{son}
M. Amirijoo, P. Frenger, F. Gunnarsson, J. Moe, K. Zetterberg, ``On self-optimization of the random access procedure in 3G Long Term Evolution,'' in \textit{Proceedings of the IFIP/IEEE International Symposium on Integrated Network Management-Workshops,} pp. 177-184, June 2009

\bibitem{sesia}
P. Bertrand and J. Jiang, ``Random Access,'' in S. Sesia, I. Toufik, and M. Baker (editors), ``LTE -- The UMTS Long Term Evolution: From Theory to Practice,'' John Wiley \& Sons, pp. 421-457, 2009

\bibitem{LTE-sim}
G. Piro, L. A. Grieco, G. Boggia, F. Capozzi, and P. Camarda, ``Simulating LTE Cellular Systems: An Open-Source Framework,'' in \textit{IEEE Transactions on Vehicular Technology,} Vol. 60, No. 2, pp. 498-513, Feb. 2011

\bibitem{SimuLTE}
A. Virdis, G. Stea, and G. Nardini, ``SimuLTE - A modular system-level simulator for LTE/LTE-A networks based on OMNeT++,'' in \textit{Proceedings of the International Conference on Simulation and Modeling Methodologies, Technologies and Applications (SIMULTECH),} pp. 59-70, Aug. 2014

\bibitem{viennaSimulator}
J. Blumenstein, J. Ikuno, J. Prokopec, and M. Rupp, ``Simulating the long term evolution uplink physical layer,'' in \textit{Proceedings ELMAR,} pp. 141-144, Sept. 2011

\bibitem{lena}
N. Baldo, M. Miozzo, M. Requena, J. Nin Guerrero, ``An Open Source Product-Oriented LTE Network Simulator based on ns-3,'' in \textit{Proceedings of the 14th ACM International Conference on Modeling, Analysis and Simulation of Wireless and Mobile Systems (MSWIM),} pp. 293-298, Nov. 2011

\bibitem{3gpp104}
3GPP, ``Base Station (BS) radio transmission and reception,'' TS 36.104, V. 13.0.0, July 2015

\bibitem{3gpp213}
3GPP, ``Physical layer procedures,'' TS 36.213, V. 12.6.0, June 2015

\bibitem{3gpp321}
3GPP, ``Medium Access Control (MAC) protocol specification,'' TS 36.321, V. 12.6.0, June 2015

\bibitem{perfeval}
F. J. López-Martínez, E. del Castillo-Sánchez, E. Martos-Naya, J. Tomás Entrambasaguas, ``Performance evaluation of preamble detectors for 3GPP-LTE physical random access channel,'' in \textit{Digital Signal Processing,} Vol. 22, Issue 3, pp. 526-534, May 2012

\end{thebibliography}
\end{document}